\documentclass[runningheads]{llncs}
\usepackage[T1]{fontenc}
\usepackage{graphicx}
\usepackage{tcolorbox}
\usepackage{hyperref}
\usepackage[dvipsnames,table]{xcolor}
\usepackage{xspace}
\usepackage[linesnumbered,ruled,vlined]{algorithm2e}
\usepackage[title]{appendix}
\usepackage{amsmath,amssymb,amsfonts}
\usepackage{subfigure}

\usepackage{multirow}
\usepackage{booktabs}
\usepackage{array}
\usepackage{caption}
\usepackage{listings}
\lstdefinestyle{vas}{
	backgroundcolor=\color{white},   
	basicstyle=\ttfamily\small,      
	keywordstyle=\color{blue},        
	commentstyle=\color{green},       
	stringstyle=\color{red},          
	numbers=left,                     
	numberstyle=\tiny\color{gray},    
	stepnumber=1,                     
	numbersep=5pt,                    
	tabsize=2,                        
	showspaces=false,                 
	showstringspaces=false,           
	breaklines=true,                  
	frame=single                      
}
\newcommand{\mx}[1]{\ensuremath{#1}\xspace}

\newcommand{\vas}{\mx{\mathcal{M}}}
\newcommand{\vasTuple}{\mx{ \langle \variableSet, \variableOrder, \transitionSet, \initialState \rangle }}

\newcommand{\variableSet}{\mx{\mathfrak{X}}}
\newcommand{\variableOrder}{\mx{\prec_{\variableSet}}}
\newcommand{\numberOfVariables}{\mx{n}}
\newcommand{\variable}[1][]{\mx{{x}_{#1}}}

\newcommand{\transitionSet}{\mx{\mathfrak{R}}}
\newcommand{\numberOfTransitions}{\mx{m}}
\newcommand{\transition}[1][]{\mx{\mathcal{R}_{#1}}}
\newcommand{\transitionTuple}{\mx{\langle \guard{i}, \update{i}, \transitionf{i} \rangle }}
\newcommand{\guard}[1]{\mx{\mathbf{g}_{#1}}}
\newcommand{\update}[1]{\mx{\mathbf{u}_{#1}}}

\newcommand{\transitionf}[1]{\mx{f_{#1}}}

\newcommand{\enabled}[2]{\mx{\texttt{enabled}(\transition[#1], \: \state[#2])}}
\newcommand{\execute}[2]{\mx{\texttt{execute}(\transition[#1], \: \state[#2])}}

\newcommand{\state}[1][]{\mx{\mathbf{s}_{#1}}}

\newcommand{\initialState}{\state[0]}

\newcommand{\upperTimeBound}{\mx{T}}

\newcommand{\probOp}{\mx{\mathsf{P}}}

\newcommand{\postUntil}{\mx{\Psi}}

\newcommand{\cslEventually}{\mx{\probOp_{=?} (\ensuremath{\Diamond^{[0, \upperTimeBound]} \, \postUntil)}}}

\newcommand{\ragtimer}{\textsc{Ragtimer}\xspace}
\newcommand{\stamina}{\textsc{Stamina}\xspace}
\newcommand{\stoolset}{\textsc{Stamina} Toolset\xspace}
\newcommand{\prism}{PRISM\xspace}
\newcommand{\storm}{Storm\xspace}
\newcommand{\modest}{Modest\xspace}
\newcommand{\modes}{\texttt{modes}\xspace}

\begin{document}

\title{\ragtimer 1.0: Rapid Rare-Event Partial State Space Construction for Stochastic VAS\\(extended version)}
\titlerunning{\ragtimer 1.0}

\author{
    Landon Taylor\inst{1}\orcidID{0000-0002-4071-3625} \and
    Joshua Jeppson\inst{1}\orcidID{0000-0002-4177-7604} \and
    Bingqing Hu\inst{1}\orcidID{0009-0006-5933-6141} \and
    Lukas Buecherl\inst{1}\orcidID{0000-0002-4844-6605} \and
    Zhen Zhang\inst{1}\orcidID{0000-0002-8269-9489}}

\authorrunning{L. Taylor et al.}

\institute{Utah State University, Utah, USA \\
\email{landon.jeffrey.taylor@usu.edu}}

\maketitle              

\begin{abstract}
Transient reachability analysis of rare events in \textit{Continuous-Time Stochastic Vector Addition Systems} (CTSVAS) such as \textit{Chemical Reaction Networks} (CRNs) has proven a formidable challenge to cutting-edge tools. Underlying a CTSVAS is a \textit{continuous-time Markov chain} (CTMC), and CTMC transient reachability analysis calls for \textit{Probabilistic Model Checking} (PMC). This analysis requires the explicit representation of a model's entire state space. 
Rare events occur with extremely low probability, compounding the challenge of probabilistic analysis. In CRNs, it is imperative to verify the probability of rare events; even a low concentration of a species can have pathological consequences.
This paper presents the \ragtimer 1.0 tool, which efficiently builds a partial state space for a CTSVAS by enumerating traces to a rare event of interest and expanding them to exploit concurrency and cycles, providing a guaranteed lower bound on  the probability of a rare event. Guaranteed lower bounds are particularly useful in synthetic biological applications because they indicate how and when a rare event can be experimentally observed. \ragtimer is an attractive alternative to existing rare event analysis methods for CTSVAS models. It outperforms existing PMC tools and refutes multiple probability estimates from rare-event stochastic simulation on multiple challenging CRN models. \ragtimer uses optimized data structures, a simple input format, and memory-safe Rust code to improve the scalability and accessibility of PMC for industry professionals.

\end{abstract}

\section{Introduction}

Transient quantitative reachability analysis of \textit{Continuous-Time Stochastic Vector Addition Systems} (CTSVAS) has proven a formidable challenge to cutting-edge tools. Underlying a CTSVAS is a \textit{continuous-time Markov chain} (CTMC), and CTMC transient reachability analysis calls for \textit{Probabilistic Model Checking} (PMC) to provide guaranteed probability results. This requires the explicit representation of a model's entire state space~\cite{Kwiatkowska_StochasticModelChecking_2007}. 
Realistic CTSVAS models, such as \textit{Chemical Reaction Networks} (CRNs) often have infinite or intractably large state spaces and are highly concurrent and cyclical. Many CRNs include rare events that occur with extremely low probability. Rare events are critical to analyze despite posing a significant verification challenge: in biochemical reaction network and synthetic biological models, even a low concentration of a species can have pathological consequences. To enable PMC for these models, the size of a state space often must be reduced to the most efficient possible degree.

This paper presents \ragtimer 1.0\footnote{The \ragtimer 1.0 artifact is available at \href{https://doi.org/10.5281/zenodo.20399349}{DOI 10.5281/zenodo.20399349}}, which efficiently and effectively builds the critical partial state space for a CTSVAS by enumerating many critical traces to states of interest and expanding those traces to exploit concurrency and cycles. \ragtimer 1.0 generates traces by analyzing dependencies, producing a partial state space that yields a guaranteed probability lower bound. It includes a novel \textit{Reinforcement Learning} (RL) trace generation approach along with data structure and input format improvements. This provides a significant improvement over the prototype of \ragtimer 0.1~\cite{Tool_Ragtimer0_1} in both runtime and probability.

\ragtimer outperforms cutting-edge PMC tools in both runtime and probability calculation, and it refutes probability estimates from cutting-edge rare-event stochastic simulation tools on multiple challenging rare-event CRN models. 
Our benchmarks show that \ragtimer is an attractive alternative to existing approaches for transient probabilistic analysis of challenging CTSVAS models. The lower probability bound provided by \ragtimer is of particular practical interest: it can allow a biologist to know the cell population required to observe a rare event, for example.
\ragtimer is the first tool in the \stoolset, which aims to improve the usability of formal methods to industry professionals.

The major contribution of this paper is the development and benchmarking of \ragtimer 1.0, including a demonstration of its scalability compared to cutting-edge tools. A minor contribution is a novel input format, which is designed to improve accessibility for users who are not experts in formal methods.

\section{Background Information}
\label{sec:background}

\ragtimer 1.0 is developed with CRNs in mind because they are challenging, real-world systems of particular interest in the field of synthetic biology~\cite{Myers_Engineeringgeneticcircuits_2010}. CRNs are represented as CTSVAS models. CTSVAS is defined as follows:

\begin{definition}[Continuous-Time Stochastic VAS]
    A CTSVAS is a tuple \vas = \vasTuple such that 
    $\variableSet = \{ \variable[i] \; | \; 0 < i \leq \numberOfVariables \}$ is the set of variables, 
    $\variableOrder$ is a total order imposed on variables in $\variableSet$, 
    $\transitionSet = \{ \transition[i] \; | \; 0 < i \leq \numberOfTransitions \}$ is the set of transitions,
    and $\initialState$ is the initial state.
    State $\state[i]$ is a vector with one non-negative integer value for each variable in \variableSet, adhering to \variableOrder.
    A transition is a tuple $\transition[i] = \transitionTuple$ such that
    $\guard{i}$ is the guard vector,
    $\update{i}$ is the update vector,
    and $\transitionf{i}$ is the transition rate function. Because CTSVAS is an extension of VAS, the guard vector exists to ensure all states are in the first orthant.
    \enabled{i}{j} indicates that in \state[j], \transition[i] may execute. Formally, $\enabled{i}{j} \triangleq \bigwedge_{0 \leq \alpha < \numberOfVariables} \guard{i}[\alpha] \leqslant \state[j][\alpha]$.
    Let \execute{i}{j} be a function returning a state that yields the result of this vector addition. Formally,
    $\state[k] = \execute{i}{j} \triangleq (\state[k] = \update{i} + \state[j] \land \enabled{i}{j})$.
    Let $p(\execute{i}{j})$ represent the single-step probability of executing \transition[i] at \state[j].
\end{definition}

Intuitively, function $\enabled{i}{j}$ ensures a transition execution cannot result in a state with negative variable assignments. Function $\execute{i}{j}$ yields state \state[k] if and only if \state[k] is equal to \update{i} + \state[j] and \transition[i] is enabled to execute in \state[j]. All values in \guard{i} are non-negative integers.

A \textit{target} is a property of interest specified in \textit{Continuous Stochastic Logic} (CSL)~\cite{Kwiatkowska_StochasticModelChecking_2007} in the format \cslEventually. Intuitively, this is read, ``What is the probability that within \upperTimeBound time units, 
the state formula \postUntil is satisfied?'' State \state is a \textit{target state} iff $\state \models \postUntil$. A target in this format requires transient analysis of the model's underlying CTMC, which requires an explicit state space representation~\cite{Tool_Storm} and can be extremely resource-intensive.

A \textit{trace} $\sigma$ is a sequence of transitions that, when fired in order, cause the target condition to be satisfied. That is, given trace $\sigma = \langle \transition[\alpha], \transition[\beta], \ldots \rangle$, it must hold that $\initialState + \sum_{\transition[i] \in \sigma} \update{i} \models \postUntil$. Let $p(\texttt{execute}({\sigma}))$ be shorthand for the product of $p(\execute{i}{j})$ for all $\transition[i] \in \sigma$, where \state[j] represents the appropriate intermediate state obtained by firing all transitions in $\sigma$ that precede \transition[i]. For example, to obtain $p(\texttt{execute}({\sigma}))$ for the 2-step trace $\sigma = \langle \transition[1], \transition[2] \rangle$, one may calculate $p(\texttt{execute}({\sigma}))$ by first calculating $p(\execute{1}{0})$ and simulating \transition[1] to obtain \state[1] = \initialState + \update{1}. Then, calculate $p(\execute{2}{1})$ and simulate \transition[2] to obtain \state[2] = \state[1] + \update{2}. Finally, $p(\texttt{execute}({\sigma})) = p(\execute{0}{1}) \times p(\execute{1}{2})$. Intuitively, $p(\texttt{execute}({\sigma}))$ represents the probability of selecting each transition during the execution of trace $\sigma$.

A \textit{rare event} is a target with extremely low probability, typically below $10^{-3}$. Rare events challenge existing approaches, especially abstraction-based approaches, because their low probability often causes them to be excluded from analysis.

An \textit{absorbing state} \state[abs] is added to a partial state space to take the place of unexplored states for probabilistic calculations. It is assumed that $\state[abs] \not\models \postUntil$, so the probability result obtained from a partial state space that contains \state[abs] is a \emph{guaranteed} lower bound on the probability from the full state space~\cite{Tool_Ragtimer0_1}. While the assumption $\state[abs] \models \postUntil$ enables the generation of a probability upper bound, the state space produced by \ragtimer is highly optimized to yield a lower bound. This extremely tight state space almost exclusively contains traces that eventually lead to a rare event. Due to the nature of rare events, these traces occur with a very low probability, so a large volume of probability (nearly 1) is redirected toward the absorbing state. Under the assumption $\state[abs] \models \postUntil$, this large (nearly 1) upper bound is not informative.

\subsection{Motivating Example}
\label{sec:motivating}
The \textit{Modified Yeast Polarization} (MYP) model~\cite{Daigle_Automatedestimationrare_2011}
has eight chemical reactions interacting over seven species. 
It was modified from the pheromone-induced G-protein cycle in \textit{Saccharomyces cerevisia}~
\cite{Drawert_Diffusivefinitestate_2010} with a constant ligand population that keeps it away from reaching equilibrium.
It has eight chemical reactions interacting with the species vector
\ensuremath{[R, L, RL, G, G_{a}, G_{bg}, G_d]} and initial state
\ensuremath{\initialState = [50,	2, 0, 50, 0, 0, 0]}.
All reaction propensities are in molecules per second.

The rare event of interest in the MYP model is the rapid build-up 
of \ensuremath{G_{bg}}: the probability of the molecule 
count of $G_{bg}$ increasing from $0$ to $50$ within $20$ seconds, specified in CSL as \ensuremath{\probOp_{=?}(\Diamond^{[0, 20]} \, G_{bg}=50)}.

\[
\begin{array}{ll}
	\transition{1}: \ \emptyset \xrightarrow{0.0038} \textrm{R}, &
	\transition{2}: \ \textrm{R} \xrightarrow{4.00\times 10^{-4}} \emptyset, \\
	\transition{3}: \ \textrm{L} + \textrm{R} \xrightarrow{0.042} \textrm{RL} 
	+ \textrm{L}, &
	\transition{4}: \ \textrm{RL} \xrightarrow{0.010} \textrm{R}, \\
	\transition{5}: \ \textrm{RL} + \textrm{G} \xrightarrow{0.011} 
	\textrm{G}_\textrm{a} + \textrm{G}_{\textrm{bg}},~~~ &
	\transition{6}: \ \textrm{G}_\textrm{a} \xrightarrow{0.100} 
	\textrm{G}_\textrm{d}, \\
	\transition{7}: \ \textrm{G}_\textrm{d} + \textrm{G}_{\textrm{bg}} 
	\xrightarrow{1.05\times 10^{3}} \textrm{G},~~~ &
	\transition{8}: \ \emptyset \xrightarrow{3.21} \textrm{RL}. \\
\end{array}
\]

Its state space is infinite due to the infinite possible values for each variable, and the rare event of interest requires at least 100 low-probability transition firings. For example, a trace may be constructed of 50 firings of \transition{8} followed by 50 firings of \transition{5}. This model incurs a large state space due to its highly concurrent and cyclical nature. It challenges cutting-edge PMC tools~\cite{Tool_CycleCommute}, and it has been the motivating model for the development of \ragtimer.

\subsection{Importance of Lower Probability Bounds}
Guaranteed lower-bound probabilities for rare events can offer practical value in many realistic use cases for \ragtimer. In synthetic biological and biochemical applications, for example, it is valuable to know how and when a rare event can be experimentally observed.
In this context, a guaranteed lower bound is needed. If the probability of a rare event occurring in a single cell is guaranteed to be at least $x$, then the probability that the event fails to occur in any one cell is at most $(1 - x)$. Across $n$ independently grown cells, the probability that the event is never observed is at most $(1 - x)^n$, which decreases as $n$ increases. To ensure that the event is observed at least once with a desired confidence level p (e.g., 95\%), one can solve for $n$ using $n \geq \log(1 - p) / \log(1 - x)$, or approximately $n \geq 3/x$ for a 95\% confidence level when $x$ is very small.
Crucially, this reasoning requires a lower bound: an upper bound on the rare-event probability would yield a lower bound on the required cell population, which could cause experimentalists to under-provision their cell culture and fail to observe the event. 

Rare events may be extremely unlikely in a single cell, but biological systems often consist of large cell populations in which such events are likely to be observed. Consequently, ignoring rare events in synthetic biology or biochemical reaction systems can have serious or even detrimental consequences, as they may manifest across a large population of cells. Providing a guaranteed lower bound on the rare-event probability enables biologists to accurately estimate the minimum cell population size needed to confidently observe the rare event of interest in wet-lab experiments.

Further, lower bounds are important when rare events represent desirable or exploitable behaviors. This distinction is relevant for biosensor design, where high-performance systems require rapid and reliable activation, particularly when detecting transient or low-abundance signals. The time required for a biosensor to reach a functional output level determines how quickly downstream processes can be triggered and constrains its practical utility. Because stochastic molecular events can cause some cells to reach target output levels much earlier than others~\cite{Ham_Stochastic_2024}, rare and early activations define the fastest detectable response in a population. In this setting, a guaranteed lower bound on the probability of early activation tells the experimentalist how many independent cells are needed to confidently observe at least one rapidly responding cell. An upper bound alone would only show that such early activation is no more likely than a given value, but it would not guarantee that the event is sufficiently likely to be observed in a finite population.

\section{Related Work}
\label{sec:related}

\ragtimer is most comparable to exact solutions for CTMC analysis. However, we also provide a brief description of its relationship to tools that find approximate solutions and abstractions.

\subsection{Probabilistic Model Checking}
For continuous-time transient reachability analysis of models with tractable state spaces, PMC is the gold standard. Modern PMC tools are powerful, efficient, and precise. However, PMC for CTMC transient analysis requires an explicit-state representation (symbolic representations such as Binary Decision Diagrams are not applicable)~\cite{Kwiatkowska_StochasticModelChecking_2007}, 
so infinite or intractably-large state spaces can exceed the computational limits of these tools. \storm~\cite{Tool_Storm}, \prism~\cite{Tool_Prism}, and \modest~\cite{Tool_Modest} are cutting-edge PMC tools for CTMC transient analysis. Many CTSVAS models pose a formidable challenge to these tools, necessitating alternative techniques.

\subsection{Approximate Solutions}
In cases where PMC is unsuccessful, approximate solutions may be generated. \textit{Statistical Model Checking} (SMC), including rare-event analysis methods, namely, \textit{Importance Sampling} and \textit{Importance Splitting}, provides approximate solutions for transient CTMC reachability. 
\textit{Importance sampling} increases the likelihood of encountering rare events during simulation by biasing the rare-event probability~\cite{Kahn_Methods_1953,Kahn_Random_1950,Roh_Statedependent_2010}. 
The \textit{weighted Stochastic Simulation Algorithm} (wSSA)~\cite{Kuwahara_Efficientexactstochastic_2008,Gillespie_Refining_2009,Daigle_Automatedestimationrare_2011} is an importance sampling algorithm that relies on a user-provided biasing scheme to estimate probabilities. Variance reduction techniques, including importance sampling, are far from being automated and motivate rare-event analysis using exact methods~\cite{Ahmadi_ComparisonWeightedStochastic_2023,Ahmadi_RareEventGuidedAnalysis_2024}. \textit{Importance splitting} uses an (often manually-computed) importance function to divide simulation traces into levels ordered by reachability probability~\cite{Rosenbluth_Monte_1955,Jegourel_Importance_2013}. 
The \modes statistical model checking tool within the \modest Toolset uses importance splitting for transient CTMC analysis~\cite{Tool_Modes,Tool_Modest}. A comparison of recent rare-event simulation methods and tools is found in~\cite{Ahmadi_ComparisonWeightedStochastic_2023}.
Alternatively, \textit{Weighted Ensemble} (WE)~\cite{Donovan_Efficientstochasticsimulation_2013,Zhang_Efficientverifiedsimulation_2007}, commonly used to analyze rare events in CRNs~\cite{Donovan_Efficientstochasticsimulation_2013,Zuckerman_WeightedEnsembleSimulation_2017}, provides a probability estimate by prioritizing promising trajectories during simulation. In contrast, our work provides a \textit{guaranteed} lower bound for the rare-event probability and requires minimal manual intervention.

\subsection{Abstraction Techniques}
\textit{Fast Adaptive Uniformization} (FAU)~\cite{Tool_FAU}, FAU+~\cite{Tool_FAUPlus}, and other sliding window abstraction techniques~\cite{Henzinger_SlidingWindowAbstraction_2009} dynamically restrict computation to states with high probability mass. They report a probability bound alongside an error bound, which characterizes the truncation error introduced by discarding low-probability states.
Similarly, SeQuaiA~\cite{Tool_SeQuaiA} is a CRN analysis tool that performs aggressive abstraction and simulation, but it does not verify a particular property~\cite{Tool_SeQuaiA}. Because these abstraction techniques are likely to abstract away low-probability transitions that lead to rare events, they are not well-suited to analyzing rare events in CTSVAS models. This limitation is empirically demonstrated by the FAU benchmarking results presented in Section~\ref{sec:benchmarks}. 

\subsection{Partial State Space Construction}
For models with infinite or intractable state spaces, it is often desirable to check only a subset of the state space. \stamina~\cite{Tool_Stamina} is an efficient state space truncation engine that produces an upper and lower probability bound for target reachability by assuming unexplored transitions lead to a target state with a probability of 0 (for a lower bound) or 1 (for an upper bound).
The prototype release of \ragtimer 0.1 \cite{Tool_Ragtimer0_1} along with its Cycle \& Commute expansion~\cite{Tool_CycleCommute} enable CTMC transient analysis for rare events beyond the limits of state-of-the-art PMC tools. It uses compositional testing to generate many traces, then expands the partial state space comprising those traces by generating parallel traces and exploring cycles from each state.

\section{Trace Generation and State Space Expansion}
\label{sec:tracegen}

\begin{figure}
	\centering
	\includegraphics[width=\linewidth]{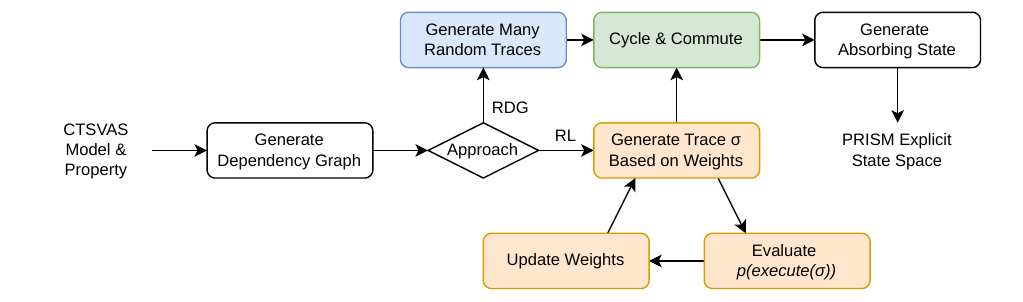}
	\caption{High-Level \ragtimer 1.0 Flowchart}
	\label{fig:flowchart}
\end{figure}

The most significant contribution of \ragtimer 1.0 is found in its efficiency and accuracy. Section~\ref{sec:benchmarks} shows dramatic improvements in both runtime and probability result compared to \ragtimer 0.1 as well as cutting-edge PMC and simulation tools. 
Algorithm~\ref{alg:ragtimer} shows a high-level view of \ragtimer 1.0, which is summarized at a high level in Figure~\ref{fig:flowchart}.

\begin{algorithm}[p]
	\caption{High-Level Overview of \ragtimer 1.0}
	\label{alg:ragtimer}
	\SetKwProg{Fn}{Function}{:}{}
	\SetKwFunction{FRagtimer}{RAGTIMER}
	\SetKwFunction{FGenerateTrace}{GENERATE\_TRACE}
	\SetKwFunction{FCommute}{COMMUTE}
	\SetKwFunction{FCycle}{CYCLE}
	\Fn{\FRagtimer{$\vas = \vasTuple$, $num\_traces$, $approach$, $\postUntil$}}{
		$DG \gets$ Dependency Graph for \vas based on \postUntil\; \label{ln:dg}
		$traces \gets \emptyset$; ~
		$state\_space \gets \emptyset$; ~
		$weights \gets \emptyset$ \;
		\If{$approach$ = Reinforcement Learning}{
			$weights \gets [$100.0 for transitions in $DG$, 1.0 for other transitions$]$\;
		}
		\While{$|\:traces\:| < num\_traces$}{
			$traces \gets traces \: \cup$ \FGenerateTrace{$\vas$, $DG$, $approach$, $weights$}\;
			\If{$approach$ = Reinforcement Learning}{
				Evaluate the trace probability and adjust $weights$ accordingly\; \label{ln:rewards}
			}
		}
		Update $state\_space$ to include states and transitions from $traces$\;
		\For{$trace \in traces$}{
			\FCommute{$\vas$, $commute\_depth$, $trace$, $state\_space$}\; 
		}
		\FCycle{$\vas$, $cycle\_length$, $state\_space$}\; 
		\Return{$state\_space$} {for explicit export to PRISM}\; \label{ln:prism} 
	}
	\Fn{\FGenerateTrace{$\vas = \vasTuple$, $DG$, $approach$, $weights$}}{
		$\state[i] \gets \initialState$; ~ 
		$trace \gets [~]$\;
		\While{$\state[i] \not\models \postUntil$}{
			\If{$approach$ = Reinforcement Learning}{
				$\transition[j] \gets$ choice weighted by $weights$ such that \enabled{j}{i}\; \label{ln:rlfire}
			}
			\Else{
				$\transition[j] \gets$ random choice from $DG$ such that \enabled{j}{i}\; \label{ln:rdg}
			}
			$\state[\alpha] \gets \execute{j}{i}$\;
			$trace$.append$(\transition[j])$\;
			$\state[i] \gets \state[\alpha]$\;
		}
		\Return{$trace$}
	}
	\Fn{\FCommute{$\vas = \vasTuple$, $depth$, $trace$, $state\_space$}}{
		$independent\_transitions \gets \{\transition[i] \; | \; \forall \state[j] \in trace, \: \enabled{i}{j} \}$ \; \label{ln:startcom}
		\For{$\transition \in independent\_transitions$}{
			Fire \transition from every state in $trace$ to get $parallel\_trace$\;
			Update $state\_space$ to include $parallel\_trace$\;
			\FCommute{$\vas$, $depth-1$, $parallel\_trace$, $state\_space$}\; \label{ln:endcom}
		}
		\If{$depth=0$}{\Return} \label{ln:depth}
	}
	\Fn{\FCycle{$\vas = \vasTuple$, $length$, $state\_space$}}{
		$cycles \gets$ permutations of multisets $Tr$ s.t. $\sum_{\transition[i] \in Tr} \update{i} = \vec{0} \land |Tr| \leqslant length$\; \label{ln:ncycles}
		\For{$c \in cycles$}{
			\For{\state[i] in $state\_space$}{
				\If{$c$ is enabled to fully execute at \state[i]}{
					Fire $c$ from $\state[i]$ and update $state\_space$ to reflect $c$\;
				}
			}
		}
		\Return
	}
\end{algorithm}

\subsection{Dependency Analysis}
The first step for \ragtimer is the creation of a static dependency graph, formalized in~\cite{Tool_Ragtimer0_1}. The dependency graph is based on the model's initial state and target, as shown in Line~\ref{ln:dg}. This static analysis step produces a directed acyclic graph containing transitions and their execution counts as vertices. Edges represent the conditions under which a transition becomes enabled to execute the prescribed number of times. For example, consider transition \transition[i], which must decrease the value of variable \variable by $k$. If $x=0$ initially, another transition (say, \transition[j]) must increase the value of \variable by $k$ to enable \transition[i]. A dependency edge is drawn $k \times \transition[i] \rightarrow l \times \transition[j]$ indicating that $k$ firings of \transition[i] are enabled by $l$ firings of \transition[j]. The graph is acyclic because cycles create a paradox: a transition cannot become enabled through its own execution. 

\begin{figure}
    \centering
    \includegraphics[width=0.5\linewidth]{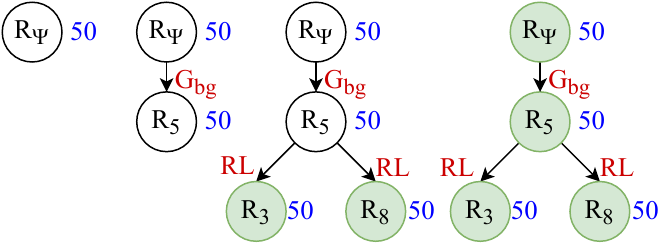}
    \caption{Dependency Graph Construction (Reproduced from \cite{Tool_Ragtimer0_1})}
    \label{fig:dg}
\end{figure}

The dependency graph describes the shortest (in number of transition firings) traces because it contains exactly the transitions that are \textit{absolutely essential} to reach a target. Figure~\ref{fig:dg} shows this process for the MYP model. The target reachability depends on \transition{5}. Before \transition{5} is able to fire, either \transition{3} or \transition {8} must fire. Thus, the dependency graph describes the need for at least 100 transitions to fire: 50 executions of \transition{5}, each preceded by a firing of \transition{3} or \transition{8}. Both \transition{3} and \transition{8} are enabled to fire in the initial state, so the graph is complete.

A naive approach to trace generation involves simply traversing the graph and firing the designated number of transitions starting with leaf nodes and moving toward the root. 
For example, if a dependency graph contains only $k \times \transition[i] \rightarrow l \times \transition[j]$, a shortest trace (length $l+k$) can be constructed with $l$ firings of \transition[j] followed by $k$ firings of \transition[i].
However, we have observed that this does not always produce reasonable traces, and the dependency graph is best utilized as a heuristic for generating high-probability traces. For example, in the MYP model, many traces may be generated with interleavings of \transition{5}, \transition{3}, and \transition{8}, but this excludes many higher-probability traces. Thus, \ragtimer 1.0 introduces the following trace generation approaches using RL and random transition firings based on the dependency graph (RDG). Figure~\ref{fig:flowchart} highlights the options of RL (via a right-bound path) and RDG (via an upward-bound path) after the generation of a dependency graph. 

\subsection{Reinforcement Learning for Trace Generation}
The problem of selecting transitions to fire while generating many traces lends itself naturally to machine learning. In addition to generating short traces based solely on dependency analysis, \ragtimer 1.0 uses RL to generate traces by assigning weights to transition firings. 
It initially assigns a high weight ($100.0$) to the transitions in the dependency graph and a small weight ($1.0$) to other transitions. This weight was determined as a heuristic through experimentation; users may wish to customize this. Enabled transitions randomly fire with respect to their weight $w$: $p(\execute{i}{k}) = w(\transition[i]) / \sum_{\transition[j] \in \{\transition[j] | \enabled{j}{k}\}} w(\transition[j])$ (see Line~\ref{ln:rlfire}).
For example, if two transitions' weights are equal, they have an equal chance of being selected, but a transition with weight $2$ has twice the chance of being chosen versus transition with weight $1$. Intuitively, the goal is to prioritize transitions that lead directly to the target, occasionally firing additional transitions to increase the diversity of traces. For example, in the MYP model, \transition{3} and \transition{8} both accomplish the goal of enabling \transition{5} to fire. However, if \transition{8} consistently had a higher $p(\execute{8}{j})$ across states \state[j], it would increase the overall trace probability and eventually have a greater weight than \transition{3}. This would cause subsequent traces to include more firings of \transition{8} and fewer of \transition{3}.

During trace generation, weights are updated to encourage firing transitions that lead to higher-probability traces, shown in Line~\ref{ln:rewards}. After each trace $\sigma$ is generated, $p(\texttt{execute}({\sigma}))$ is calculated, and a rolling average over the most recently generated traces is maintained. If $p(\texttt{execute}({\sigma}))$ exceeds the rolling average, the weight increases for each transition in the trace. Similarly, if $p(\texttt{execute}({\sigma}))$ is lower than the rolling average, the weight decreases for each transition in the trace. Weights increase or decrease proportional to the number of transition firings. For example, if a transition fires $x$ times in a trace that exceeds the rolling average, its weight is increased $x$ times. Section~\ref{sec:benchmarks} shows that this approach greatly outperforms \ragtimer 0.1 in terms of time and memory usage, and it generates a partial state space with significantly improved results. 

\subsection{Random Dependency Graph for Trace Generation}
\ragtimer 0.1 uses compositional testing implemented in the third-party tool, IVy~\cite{Tool_IVy}, translates IVy into C++, then compiles and executes the C++ code to randomly generate traces~\cite{Tool_Ragtimer0_1}. To eliminate the heavy burden of translating and then compiling the code, \ragtimer 1.0 generates traces entirely in Rust. It randomly chooses enabled transitions from the dependency graph (Line~\ref{ln:rdg}) to construct traces that, on average, advance toward a target. This random approach is favored over deterministic approaches because it is able to quickly and uniformly sample a state space. Further, the compositional testing approach used by IVy in \ragtimer 0.1 frequently generates duplicate traces, which contributes a significant burden to runtime and memory. 
This approach can quickly generate a large set of unique traces, and it sees a massive speedup compared to \ragtimer 0.1. It also improves accessibility: there is less reliance on third-party tools, and the faster runtime enables execution on lower-grade equipment.

\subsection{Cycle \& Commute}
A significant enhancement to \ragtimer 0.1 was \textit{Cycle \& Commute}, which finds concurrency and cyclical behavior at the model level and automatically generates additional traces that contain this behavior~\cite{Tool_CycleCommute}. If a transition is enabled to fire in every state, it is possible to increase the probability lower bound by firing that transition at every state along a trace. This produces a parallel trace (Lines~\ref{ln:startcom}-\ref{ln:endcom}). This takes advantage of the highly concurrent nature of many CTSVAS models by aggressively creating parallel and interleaving traces, contrary to the conventional approach of reducing interleavings. 

For example, in the MYP model, firing \transition{1} does not impact the ability of \transition{3}, \transition{5}, or \transition{8} to fire. Thus, a parallel trace may be generated by firing \transition{1} once before firing the sequence of transitions from a trace composed of \transition{3}, \transition{5}, or \transition{8}. Because an increase to the value of species $\mathrm{R}$ increases the rate of \transition{3}, the parallel trace including a firing of \transition{1} occurs with a higher probability than the original trace. \ragtimer exploits this by scaling it: up to a specified recursion depth, it adds parallel traces everywhere they are possible.

Further, adding cycles into the explicit state space redirects a portion of the probability away from an absorbing state and back toward a target, increasing the probability lower bound~\cite{Tool_CycleCommute}. Cycles are added by enumerating multisets of transitions $Tr$ such that $\sum_{\transition[i] \in Tr} \update{i} = \vec{0}$. Section~\ref{sec:benchmarks} shows a massive improvement in probability lower bounds when Cycle \& Commute is executed. For example, the cycle $\transition{1} \rightarrow \transition{2} \rightarrow \transition{1} \ldots$ is enabled to fire at every state in the MYP model. While the probability of this cycle is typically directed to an absorbing state and assumed to never reach a target, this allows the probability from this cycle to be recaptured and considered toward target reachability. Without explicit termination conditions, this process may never terminate due to CRN models' infinite state spaces. Thus, termination relies on explicit maximum lengths of cycles (Line~\ref{ln:ncycles}) and commuting depth (Line~\ref{ln:depth}). While trace generation in \ragtimer is random, parallel traces and cycles are added deterministically.

\section{Implementation Details}
\label{sec:implementation}

This section describes the implementation of \ragtimer 1.0, including a novel custom input format, its inclusion in the \stoolset, an overview of its algorithm, and engineering improvements over its previous implementation. 

\subsection{CTSVAS Input Language}
To improve accessibility for synthetic biologists and other industry professionals, \ragtimer 1.0 takes as input a novel, straightforward input format. This format is designed to represent generic CTSVAS models as well as CRNs. It is described in detail in Appendix~\ref{appendix:language}. However, to demonstrate the simplicity of the input language, Figure~\ref{lst:simple} shows a simple specification in which two transitions interact over three variables, and the property of interest inquires about the probability of increasing the value of \texttt{S3} from 300 to 400. The input format is designed to closely resemble typical CRN notation, demonstrated in Section~\ref{sec:motivating}. For example, Figure~\ref{lst:simple} notates the transition $\texttt{R1}: \ \texttt{S1} \xrightarrow{0.4} 2\texttt{S2}$ by describing the transition name, then notating that \texttt{S1} decreases by 1, \texttt{S2} increases by 2, and the transition occurs with a reaction rate constant of 0.4. In a CRN, the transition rate is defined as a function of this reaction rate constant~\cite{Myers_Engineeringgeneticcircuits_2010}. In this model, because species \texttt{S1} is consumed in transition \texttt{R1}, the rate formula $f_1(\state[i]) = 0.4(\state[i][\texttt{S1}])$.
For a realistic model, Figure~\ref{lst:mypinput} in Appendix~\ref{appendix:language} shows an specification of the motivating example, the MYP model. We argue that a simple custom input format is justified for \ragtimer, as it improves the accessibility of our tool to industry professionals without a background in formal modeling and verification.

\begin{figure}[htbp]
	\centering
	\begin{lstlisting}[style=vas]
var S1 init 100
var S2 init 200
var S3 init 300
target S3 = 400
transition R1
	decrease S1 1
	increase S2 2
	const 0.4
transition R2
	decrease S1 1
	increase S3 1
	const 0.3
	\end{lstlisting}
	\caption{Simple CTSVAS Specification}
	\label{lst:simple}
\end{figure}

\subsection{Probabilistic Analysis}
Once \ragtimer has finished constructing a partial state space, it exports the state space to the \prism explicit model format (Line~\ref{ln:prism}). This format represents states and transitions in plain text and is widely cross-compatible: it can easily be checked in mainstream PMC tools such as \prism and \storm.

\subsection{Efficiency Improvements over \ragtimer 0.1}
To store explicit states and traces, we utilize custom prefix trees in place of traditional hash maps. To our knowledge, the use of prefix trees for explicit state and trace representation is novel, and it greatly improves the memory efficiency and maintains the time complexity of our tool. During trace generation, transition sequences are stored in prefix trees to quickly check if a trace already exists. States are organized in a prefix tree as described in~\cite{Taylor_Prefix_2025} to quickly check if a state exists while improving memory usage for storing the explicit state space.

\ragtimer 0.1 utilizes the \prism API for probability calculation, which provided a balance between convenience and efficiency for transition rate calculation. Now, \ragtimer 1.0 calculates rates and probabilities internally to improve efficiency. 
Further, \ragtimer 0.1 relies on assume-guarantee reasoning within the IVy tool. This requires a considerable amount of time and memory. \ragtimer 1.0 replaces this approach with methods that have proven more effective: transition selection based on RL and the dependency graph. 
Section~\ref{sec:benchmarks} shows the payoff of these decisions.

\subsection{The \stoolset}
\ragtimer 1.0 is included in the preliminary release of the \stoolset~\cite{Tool_StaminaToolset}, a cross-platform compatible, open-source toolset that will eventually also include \stamina~\cite{Tool_Stamina}, Wayfarer~\cite{Tool_Wayfarer}, and several visualization and analysis tools.
The \stoolset is the first PMC toolset built in Rust, and it is designed to be modular, efficient, and accessible to non-technical users. 
The \stoolset will feature a graphical interface to improve accessibility.
\section{Benchmarks}
\label{sec:benchmarks}

\ragtimer 1.0 has been tested on a set of realistic rare-event CRN benchmarks. This section describes our set of models, experimental setup, our results, and a comparison to existing state-of-the-art tools.

\subsection{Rare-Event CRN Models}
We benchmarked \ragtimer 1.0 using the same set of realistic CRN models used to benchmark \ragtimer 0.1 to enable a direct comparison. These models are described below, with an extended description available in Appendix~\ref{appendix:models}. Each species in these models is an unbounded, non-negative integer. In early design stages of biochemical reaction and synthetic biological systems, it is difficult to accurately bound species count in a model, especially when the property of interest is a rare event and inadequately applying a manual, tight bound may erroneously lead to the reduced likelihood or even removal of the rare event probability. These models are preferred for this task over other benchmark sets because they present exactly the challenges \ragtimer was developed to tackle: they are highly-concurrent, highly-cyclical CRNs with large or infinite state spaces and rare events. These models are commonly used to benchmark rare event simulators. Despite their apparent simplicity, these models push or exceed the limits of cutting-edge tools. Because of the long time span required for probabilistic analysis of these models, we prioritized these four realistic models taken from case studies in biochemical reaction networks and synthetic biology.

The \textit{Modified Yeast Polarization} (MYP) model~\cite{Daigle_Automatedestimationrare_2011} 
is the motivating example described in Section~\ref{sec:motivating}.

The \textit{Enzymatic Futile Cycle} (EFC) model~\cite{Kuwahara_Efficientexactstochastic_2008} consists of six reactions acting on six species. This model requires 25 firings of each of two reactions to reach the rare-event target, which encodes a decrease of a species from a count of 50 to a count of 25.
This target of a \textit{decreased} variable value differentiates this model from the others. Its dependency graph is based on this decrease. Because it has exactly one shortest trace and it is highly concurrent and cyclical, it challenges existing tools but is well-suited for \ragtimer. 

The \textit{Simplified Motility Regulation} (SMR) model~\cite{Kearns_Cellpopulationheterogeneity_2005} consists of nine species reacting through twelve reactions. Its rare-event target is a query on the probability that a species increases from a count of 10 to a count of 20. This increase requires several well-timed firings of multiple low-probability dependency transitions, so it is challenging to observe. It is highly concurrent and cyclical, and it includes reactions that compete with high-probability reactions that lead to the rare-event target, so it is well-suited for analysis with \ragtimer.

Finally, the \textit{Single-Species Production Degradation} (SSPD) model~\cite{Kuwahara_Efficientexactstochastic_2008} is a simple production-degradation interaction of two reactions over two species. The rare event of interest is the increase of a species from a count of 40 to a count of 80, which requires many low-probability reaction firings. This model demonstrates the effectiveness of \ragtimer 1.0 on simple models. 

\subsection{Experiment Setup}
The following results were obtained on an AMD Ryzen Threadripper 12-Core 3.5 GHz Processor and 132 GB of RAM, running Ubuntu 22.04 LTS. We allocated one CPU for each test and compared to other cutting-edge tools. Unless otherwise mentioned, all models are \emph{unbounded}.

\subsection{Results}
\begin{figure}[tbh]
	\centering
	\includegraphics[width=\linewidth]{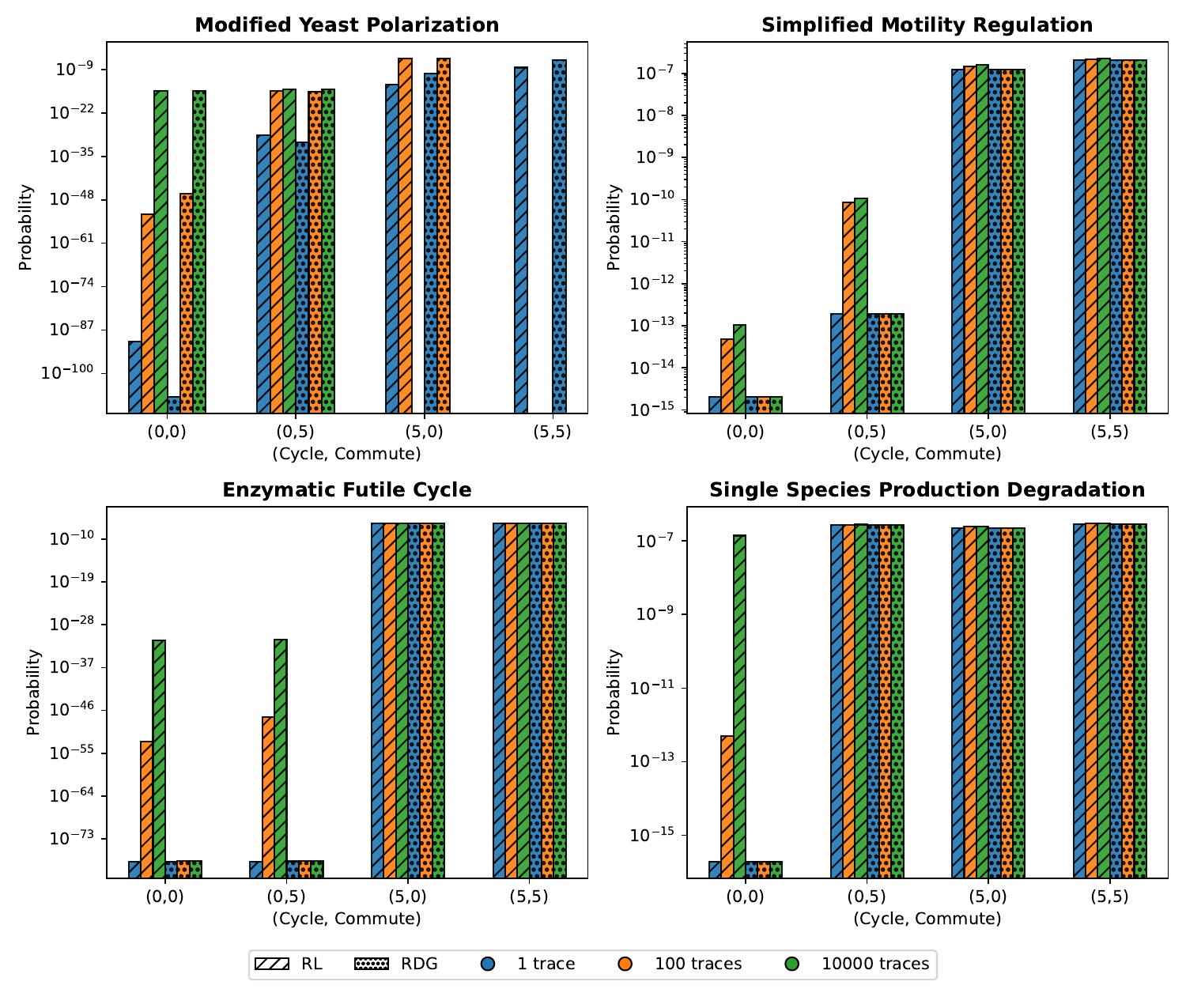}
	\caption{Graphical Summary of Results from \ragtimer 1.0.}
	\label{fig:results}
\end{figure}

We benchmarked various configurations of parameters for Ragtimer 1.0 on each of the four CRN models. Tables \ref{tab:myp}, \ref{tab:efc}, \ref{tab:smr}, and \ref{tab:sspd} in Appendix~\ref{appendix:results} show the full results, and a summary is provided in Figure~\ref{fig:results}.
Each plot contains results from a single model. Within each plot, the x-axis shows the pairs (maximum cycle length, maximum commute recursion depth), and the y-axis shows the probability result. This probability is a lower bound on the true rare-event probability. The color of the bar indicates the number of traces generated \textit{before} running Cycle \& Commute. Striped bars represent traces generated with RL, while dotted bars represent traces generated with RDG. An absence of bars indicates a timeout.

As shown in Figure~\ref{fig:results}, as cycles and parallel traces are added to each model, the probability lower bound approaches the true probability value. A user may wonder what parameters are good enough to give a reasonable probability estimate. For our benchmark set, the pair (5, 5) is sufficient to reach a plateau and approach the true probability, so this is a reasonable starting point. If the probability value does not meet a user's expectations with the pair (5, 5), a user may increase these values until a plateau is reached. Our results show that each model benefits from unique parameter modifications. For example, the MYP model benefits from cycles and parallel traces alike, but the EFC model benefits almost exclusively from cycles. We encourage users to test variations in parameters to find successful configurations. However, an unexperienced user can obtain a successful result by selecting pairs of equal values. Because \ragtimer often requires only seconds to execute, there is a low cost for trying parameter combinations. This provides a reasonable tradeoff in probability value versus the additional runtime and memory requirements from higher parameters. Further, we observe that \ragtimer 1.0 typically produces higher probability results when using RL to generate traces compared to RDG. This suggests that RL is an effective tool for quickly identifying the traces that make up a large portion of the rare event's probability. 

\subsection{Comparison to Existing Tools}

Each model highlights unique attributes of \ragtimer. For example, the MYP model highlights the scalability of this approach. 
Because of the efficiency improvements in \ragtimer 1.0, we now observe a drastic improvement: it generates 100 traces with RL and finds a probability of 1.67E-06 within  minutes---\emph{over 30\%} greater than the published best-case estimates from wSSA and WE (1.24E-06 and 1.48E-6, respectively)~\cite{Daigle_Automatedestimationrare_2011,Donovan_Efficientstochasticsimulation_2013} and within the range reported by \textsc{Stamina}, ([1.43E-06, 1.49E-05]), 
which took 4 hours to compute.
Further, this state space requires only 40 MB of memory, allowing for analysis without specialized computational resources. Because the model is highly concurrent and cyclical, the approach benefits greatly from the addition of cycles and parallel traces.
As shown in Figure~\ref{fig:results}, the greatest benefit is found from executing Cycle \& Commute after generating many traces. For example, the highest probability achieved through trace generation alone is 4.34E-16, while the greatest probability achieved after Cycle \& Commute is 1.67E-06, ten orders of magnitude greater.

The EFC model highlights unique challenges. This model has exactly one shortest trace, so it benefits from RL-based trace generation that prioritizes short, diverse traces with relatively high probabilities. Even when only 1 trace is generated, adding cycles results in a probability of 1.72E-07, nearly equal to the true probability of 1.74E-07~\cite{Kuwahara_Efficientexactstochastic_2008}, in about 3 milliseconds. We also observe that some benefit is obtained by generating a large set of traces (10,000 traces generated by RL reaches a probability of 4.50E-32), but the addition of cycles provides the greatest increase to the probability. We theorize that this is because a large probability mass is able to be redirected back into the partial state space (and thus, toward a target state) rather than to an absorbing (assumed non-target) state. 

The SMR model similarly benefits from the addition of cycles. Despite the size and complexity of the model, \ragtimer quickly reaches a probability of 2.21E-07 using RL trace generation plus Cycle \& Commute, \textit{which refutes the estimate of 2.16E-07 from wSSA}~\cite{Ahmadi_ComparisonWeightedStochastic_2023}. Despite an infinite state space, \ragtimer required only 400 MB of memory to reach this result.

Finally, the SSPD model (shown in the bottom right of Figure~\ref{fig:results}) is the simplest model we test. We observe that within 1 second, \ragtimer returns a probability of 2.95E-07, which is equal to the probability found through wSSA~\cite{Kuwahara_Efficientexactstochastic_2008}.

Table~\ref{tab:results} summarizes a comparison of \ragtimer 1.0 to its prototype release (\ragtimer 0.1) and four other tools: \storm~\cite{Tool_Storm}, \stamina~\cite{Tool_Stamina}, a recent \textit{Bounded Model Checking} (BMC) tool~\cite{Ahmadi_RareEventGuidedAnalysis_2024}, and \prism with FAU~\cite{Tool_FAU} enabled. 
We report a probability result, state count, and runtime in milliseconds for each tool on each model. For most tools, $P_{min}$ indicates the calculated probability lower bound. \storm typically computes an exact probability, but because our theoretically-unbounded models were bounded to improve \storm's ability to check them, the resulting probability is a lower bound. Each variable was bounded to $[0,250]$. For \prism with FAU, $P_{est}$ represents the reported probability estimate and $P_{loss}$ represents the reported probability loss (i.e., the error estimated for $P_{est}$). \prism with FAU reports unique results because it relies on abstraction to produce estimates rather than guaranteeing probability lower bounds. Each variable for \prism with FAU was bounded to $[0,1000]$.
\ragtimer 1.0 consistently outperforms or is competitive against state-of-the-art probabilistic verification approaches, making transient analysis of CRNs feasible to users without specialized resources. Especially on the more complex MYP and SMR models, \ragtimer 1.0 provides an attractive alternative, especially when time and memory are limited. On the less complex models, \ragtimer 1.0 remains competitive with existing methods, reaching a similar probability result in a similar amount of time. While \ragtimer 1.0 does not always outperform other approaches in every metric, it offers the best balance between probability, runtime, and memory usage.

\begin{table}[tbh]
	\centering
	\caption{Summary of Result Comparisons}
	\begin{tabular}{|p{2.9cm}|l|l|l|l|l|}
		\hline 
		\multicolumn{2}{|l|}{} & \textbf{MYP} & \textbf{EFC} & \textbf{SMR} & \textbf{SSPD} \\ \hline
		\multirow{2}{*}{\textbf{Storm}} & $P_{min}$   &  & 1.74E-07 &  & 2.99E-07 \\ 
		\multirow{2}{*}{\textbf{(bounded)}} & states & Timeout & 2.98E+02 & Memout & 8.10E+01 \\ 
		& ms     &  & 1.00E+02 &  & 3.00E+01 \\ \hline
		
		\multirow{3}{*}{\textbf{Stamina}} & $P_{min}$   &  & 0.00E+00  & 2.41E-07 & 0.00E+00 \\ 
		& states & Timeout & 8.90E+01  & 8.12E+06 & 8.90E+01 \\ 
		& ms     &  & 1.20E+03  & 2.77E+06 & 9.50E+02 \\ \hline
		
		\multirow{2}{*}{\textbf{BMC}} & $P_{min}$   & 9.08E-07 & 1.59E-07 & 3.48E-08 &  \\ 
		\multirow{2}{*}{\textbf{(5\% threshold)}} & states & 2.25E+06 & 4.18E+02 & 3.19E+05 & Error \\ 
		& ms     & 2.35E+06 & 8.00E+02 & 2.40E+03 &  \\ \hline
		
		& $P_{est}$   & 0.00E+00 & 1.71E-07 & 1.11E-04 & 2.98E-07 \\ 
		\textbf{\prism FAU} & $P_{loss}$  & 3.59E-03 & 1.99E-06 & 9.48E-06 & 1.52E-06 \\ 
		\textbf{(bounded)} & states & 8.01E+04 & 2.14E+02 & 6.86E+06 & 7.60E+01 \\ 
		& ms     & 3.69E+07 & 1.28E+03 & 2.46E+05 & 1.29E+03 \\ \hline
		
		\multirow{2}{*}{\textbf{\ragtimer 0.1}} & $P_{min}$   & 5.52E-27 & 4.38E-10 & 7.51E-14 & 2.99E-07 \\ 
		\multirow{2}{*}{\textbf{(best config)}} & states & 2.34E+04 & 5.20E+01 & 4.22E+04 & 8.50E+01 \\ 
		& ms     & 6.68E+04 & 7.35E+05 & 2.31E+04 & 2.12E+04 \\ \hline
		
		\multirow{2}{*}{\textbf{\ragtimer 1.0}} & $P_{min}$   & 1.67E-06 & 1.72E-07 & 2.21E-07 & 2.95E-07 \\ 
		\multirow{2}{*}{\textbf{(best config)}} & states & 2.96E+05 & 1.09E+02 & 4.87E+05 & 5.90E+01 \\ 
		& ms     & 1.00E+06 & 6.00E+00 & 4.17E+06 & 6.00E+01 \\ \hline
	\end{tabular}
	\label{tab:results}
\end{table}

While a cutting-edge PMC tool like Storm is the gold standard for CTMC transient reachability analysis, we find that rare events in CRNs push it past its limits. We tested each model in \storm with the \textsc{Sylvan} library~\cite{vanDijk_Sylvanmulticoreframework_2017}. \storm exceeded the 2-hour time limit for the MYP model and the 128 GB memory limit for the SMR model. In contrast, \ragtimer 1.0 was able to compute a highly accurate probability within the time and memory limit on all four models.

\stamina, a state space truncation tool that efficiently computes a probability range~\cite{Tool_Stamina}, is also unable to calculate a probability for the MYP model within the 2-hour time limit. Further, \ragtimer 1.0 produces a significantly tighter lower probability bound in comparison to \stamina on the EFC and SSPD models.

The BMC tool\footnote{Obtained from \url{https://github.com/fluentverification/bmc_counterexample}} presented in~\cite{Ahmadi_RareEventGuidedAnalysis_2024} provides reasonable probability bounds for three models, though we are unable to reproduce the results as they appear in the paper. The BMC tool reports incrementing integer thresholds, but the paper reports probability thresholds. 
Moreover, \ragtimer 1.0 reliably produces a higher probability result in a shorter runtime. Further, we observe an error in the tool when checking the SSPD model. Due to the nature of BMC, this tool relies on a threshold for diminishing returns. To obtain the results presented in Table~\ref{tab:results}, the tool terminates when the probability result does not improve by at least 1\% in one pass.

\prism with FAU requires a bounded model, so $P_{est}$ is an \textit{estimate} of the probability \textit{lower bound}. For every model, we chose the reasonably high bounds of $[0,1000]$ for each variable. While FAU is a promising approach for many CTMCs,
it does not produce reliable rare-event probabilities. The probability loss is often closer to the rare-event probabilities than the probability estimate. In CRNs, for example, it is often desirable for a biologist to understand the cell population required to observe an event. 
We speculate that the powerful abstraction techniques in FAU often ignore rare events altogether. In comparison, \ragtimer is able to find the portions of a state space that efficiently contribute to the rare-event probability, making it a more reliable tool for transient analysis of rare events in CTSVAS.

\ragtimer 1.0 performs particularly well in comparison to \ragtimer 0.1. While \ragtimer 0.1 was able to quickly generate traces, these traces' cumulative probability was often negligibly low. In the best case, the first implementation of Cycle \& Commute increased the probability to 5.26E-26. Within comparable runtimes and memory usage, \ragtimer 1.0 improves each probability result by several orders of magnitude compared to \ragtimer 0.1.

Because \ragtimer provides a \textit{guaranteed} result, we do not provide a detailed benchmark of tools that provide approximate results. However, we have evaluated several CRN models against approximate methods as follows. The \modes tool from the \modest Toolset efficiently estimates rare-event probabilities in the presented case studies, with results closely aligning with those reported by \cite{Kuwahara_Efficientexactstochastic_2008}, \cite{Daigle_Automatedestimationrare_2011}, and \cite{Donovan_Efficientstochasticsimulation_2013}.
However, \modes depends on a compositional importance function for rare-event simulation. While it is feasible for users to alter the importance function, such modifications necessitate user involvement and a thorough understanding of both the input model and the \modest language. As reported in~\cite{Tool_Ragtimer0_1}, \modes requires a compositional importance function for rare-event simulation, which limits the use of global variables and requires significant user intervention.

Other simulation methods like wSSA and WE require advance knowledge to approach a true probability value. Additionally, \ragtimer has demonstrated that it significantly outperforms their best-case estimates. For the MYP model, wSSA reported a best-case probability estimate of 1.24E-06 in about 13 hours~\cite{Daigle_Automatedestimationrare_2011}, exceeded by 34\% by the lower bound of 1.67E-06 discovered by \ragtimer. The best-case estimate from WE is 1.48E-06~\cite{Donovan_Efficientstochasticsimulation_2013}, exceeded by 13\% by the lower bound from \ragtimer 1.0. To the best of our knowledge, these are the most recent results reported for wSSA and WE on the MYP model.
Because \ragtimer 1.0 is grounded in formal methods, its improved lower bound calculation refutes the results found from simulation and more accurately calculates this rare-event probability. Additionally, as shown above, \textit{\ragtimer 1.0 produces guaranteed probability lower bounds that refute or compete with those obtained by rare-event simulation for each model}. 

\ragtimer improves the accuracy of lower bounds in several cases, which is valuable to experimental analysis. While its state space construction is random in nature and thus not guaranteed to be optimal, they are shown to be effective compared to deterministic methods. For example, by improving the lower bound compared to wSSA by 34\%, biologists can potentially save 34\% of the cell population (and associated costs) in their efforts to observe the rare event. This bolsters the practical usability of \ragtimer by industry professionals.

\ragtimer enables analysis where cutting-edge tools currently face roadblocks. Because \ragtimer can outperform rare-event simulation methods in both probability accuracy and runtime, it provides an attractive alternative to existing tools and techniques. 

\section{Conclusion}
\label{sec:conclusion}

\ragtimer 1.0 is a highly-efficient trace generation and partial state space construction tool for the transient analysis of rare events in CTSVAS.
It makes significant advancements in transient rare-event reachability analysis for CTSVAS models. It efficiently constructs a streamlined partial state space that takes advantage of concurrent and cyclical behavior to compute a guaranteed probability lower bound on a property of interest. 

Our evaluation on a set of real-world CRN models demonstrates that \ragtimer is a strong alternative for CTSVAS analysis. \ragtimer 1.0 not only significantly outperforms its previous prototype release; it outperforms and produces results that are competitive against cutting-edge tools, and it refutes the probability estimates generated by rare-event simulation techniques. 

As the first tool in the \stamina Toolset, \ragtimer enhances the practicality and accessibility of formal techniques to industry professionals, particularly in the field of synthetic biology. Future work includes additional tool development and integration in the \stamina Toolset including the acceptance of a broader range of custom rate functions, as well as proofs of correctness within \ragtimer's codebase.

\section*{Acknowledgment}
L.T., J.J., and B.H. are supported by the National Science Foundation under Grant Nos. 2422206 and 1856733. L.T. and J.J. are also partially supported by the University of Utah Center for Medical Cannabis Research (CMCR) Seed Grant. B.H. is partially supported by the Utah State University Integrated Team Research Grant. Any opinions, findings, and conclusions or recommendations expressed in this material are those of the authors and do not necessarily reflect the views of the funding agencies.

%
%
\bibliographystyle{splncs04}
\bibliography{assets/refs,assets/tools}


\newpage 

\begin{subappendices}
	\renewcommand{\thesection}{\Alph{section}}%

\section{CRN Models}
\label{appendix:models}

\subsection{Modified Yeast Polarization}

The \emph{Modified Yeast Polarization} model~\cite{Daigle_Automatedestimationrare_2011} 
was modified from the pheromone-induced G-protein cycle in Saccharomyces cerevisia 
\cite{Drawert_Diffusivefinitestate_2010} with a constant ligand population that keeps it away from reaching equilibrium.
It has eight chemical reactions interacting with the species vector
\ensuremath{[R, L, RL, G, G_{a}, G_{bg}, G_d]} and initial state
\ensuremath{\initialState = [50,	2, 0, 50, 0, 0, 0]} .
All reaction propensities are in molecules per second.
The rare event of interest for our motivating example is the rapid build-up 
of \ensuremath{G_{bg}}: the probability of the molecule 
count of $G_{bg}$ increasing from $0$ to $50$ within $20$ seconds, specified in CSL as \ensuremath{\probOp_{=?}(\Diamond^{[0, 20]} \, G_{bg}=50)}. This model is highly-concurrent and challenges cutting-edge 
tools~\cite{Tool_CycleCommute}. 

\[
\begin{array}{ll}
	\transition{1}: \ \emptyset \xrightarrow{0.0038} \textrm{R}, &
	\transition{2}: \ \textrm{R} \xrightarrow{4.00\times 10^{-4}} \emptyset, \\
	\transition{3}: \ \textrm{L} + \textrm{R} \xrightarrow{0.042} \textrm{RL} 
	+ \textrm{L}, &
	\transition{4}: \ \textrm{RL} \xrightarrow{0.010} \textrm{R}, \\
	\transition{5}: \ \textrm{RL} + \textrm{G} \xrightarrow{0.011} 
	\textrm{G}_\textrm{a} + \textrm{G}_{\textrm{bg}},~~~ &
	\transition{6}: \ \textrm{G}_\textrm{a} \xrightarrow{0.100} 
	\textrm{G}_\textrm{d}, \\
	\transition{7}: \ \textrm{G}_\textrm{d} + \textrm{G}_{\textrm{bg}} 
	\xrightarrow{1.05\times 10^{3}} \textrm{G},~~~ &
	\transition{8}: \ \emptyset \xrightarrow{3.21} \textrm{RL}. \\
\end{array}
\]

\subsection{Enzymatic Futile Cycle}

This CRN models a futile cycle with six reactions acting on six species~\cite{Kuwahara_Efficientexactstochastic_2008}, defined by vector 
\ensuremath{[S_1, S_2, S_3, S_4, S_5, S_6]} 
with the initial state \ensuremath{\initialState = [1, 50, 0, 1, 50, 0]}. The rare event of interest is 
\ensuremath{\probOp_{=?}(\Diamond^{[0, 100]} \, S_5 = 25)}, or a query of the probability that within 100 seconds, $S_5$ decreases from $50$ to $25$. This model is concurrent and cyclical, and it challenges cutting-edge 
tools~\cite{Tool_CycleCommute}. 

\[
\begin{array}{ll}
	\transition{1} : \ S_1 + S_2 \xrightarrow{1.0} S_3, &
	\transition{2} : \ S_3 \xrightarrow{1.0} S_1 + S_2,~~~ \\
	\transition{3} : \ S_3 \xrightarrow{0.1} S_1 + S_5, &
	\transition{4} : \ S_4 + S_5 \xrightarrow{1.0} S_6,~~~ \\
	\transition{5} : \ S_6 \xrightarrow{1.0} S_4 + S_5,~~~ &
	\transition{6} : \ S_6 \xrightarrow{0.1} S_4 + S_2.
\end{array}
\]

\subsection{Simplified Motility Regulation}

This model represents the genetic mechanism which regulates flagella formation in \emph{Bacillus subtilis} and consists of consists of nine species and twelve reactions~\cite{Kearns_Cellpopulationheterogeneity_2005}. The initial molecule count for species vector 
[codY, flache, SigD\_hag, CodY, CodY\_flache, hag,CodY\_hag, 
SigD, Hag] forms the initial state 
\ensuremath{\initialState = [1, 1, 1, 10, 1, 1, 1, 10, 10]}. 
The rare event property is 	\ensuremath{\probOp_{=?}(\Diamond^{[0, 10]} \, \textrm{CodY} = 20)}, or a query on the probability that within 10 seconds, CodY increases from 10 to 20.
\[
\begin{array}{l}
	\transition{1} : \ \textrm{codY}  \xrightarrow{0.1} \textrm{codY} + 
	\textrm{CodY}, \\
	\transition{2} : \ \textrm{CodY} \xrightarrow{0.0002} \emptyset ,\\
	\transition{3} : \ \textrm{flache} \xrightarrow{1.0} \textrm{flache} + 
	\textrm{SigD}, ~~~ \\
	\transition{4} : \ \textrm{SigD} \xrightarrow{0.0002} \emptyset,\\
	\transition{5} : \ \textrm{SigD\_hag} \xrightarrow{1.0} \textrm{SigD} + 
	\textrm{hag} + \textrm{Hag},~~~ \\
	\transition{6} : \ \textrm{Hag} \xrightarrow{0.0002} \emptyset,\\
	\transition{7} : \ \textrm{SigD} + \textrm{hag}\xrightarrow{0.01} 
	\textrm{SigD\_hag},~~~ \\
	\transition{8} : \ \textrm{SigD\_hag} \xrightarrow{0.1} \textrm{SigD} + 
	\textrm{hag},\\
	\transition{9} : \ \textrm{CodY} + \textrm{flache} \xrightarrow{0.02} 
	\textrm{CodY\_flache},~~~ \\
	\transition{10} :  \textrm{CodY\_flache} \xrightarrow{0.1} \textrm{CodY} 
	+ \textrm{flache},\\
	\transition{11} :  \textrm{CodY} + \textrm{hag} \xrightarrow{0.01} 
	\textrm{CodY\_hag},~~~ \\
	\transition{12} :  \textrm{CodY\_hag} \xrightarrow{0.1} \textrm{CodY} + 
	\textrm{hag}.\\
\end{array}
\]

\subsection{Single Species Production-Degradation Model}
The model describes a production-degradation 
interaction between two species \cite{Kuwahara_Efficientexactstochastic_2008}. The initial state for the species vector \ensuremath{[S_1, S_2]} is 
\ensuremath{\initialState = [1, 40]}, and the desired CSL property is 
\ensuremath{\probOp_{=?}(\Diamond^{[0, 100]} \, S_2 = 80)}, or a query on the probability that within 100 seconds, the count of species $S_2$ increases from $40$ to $80$.
\[
\begin{array}{ll}
	\transition{1} : S_1 \xrightarrow{1.0} S_1 + S_2 ~~~&~~~
	\transition{2} : S_2 \xrightarrow{0.025} \emptyset
\end{array}
\]

\newpage

\section{Full Result Tables}
\label{appendix:results}

Tables \ref{tab:myp}, \ref{tab:efc}, \ref{tab:smr}, and \ref{tab:sspd} include the full data summarized in Figure~\ref{fig:results}. In each table, the ``Approach'' column indicates if traces were generated using Reinforcement Learning (RL) or a random firing from transitions in the dependency graph (RDG). ``Traces'' shows the number of traces generated. ``Cycles'' shows the maximum length of cycles added to the state space after trace generation. ``Commute'' shows the maximum recursion depth for commuting transitions. ``Probability'' shows the probability outcome when the explicit partial state space is passed to the \prism model checker. This probability is a lower bound on the true rare event probability. ``Time (ms)'' shows the number of milliseconds required to generate the partial state space, and ``Bytes'' shows the total memory required for partial state space construction.

\begin{table}[p]
	\caption{Modified Yeast Polarization}
	\label{tab:myp}
	\resizebox{\textwidth}{!}{%
		\begin{tabular}{lllllll}
			\hline
			\textbf{Approach} & \textbf{Traces} & \textbf{Cycle} & \textbf{Commute} & \textbf{Probability} & \textbf{Time (ms)} & \textbf{Bytes} \\ \hline
			RL                & 1               & 0              & 0                & 3.23E-91             & 1.00E+00               & 3.93E+05                  \\
			RL                & 1               & 0              & 5                & 2.00E-29             & 1.45E+02               & 5.31E+06                  \\
			RL                & 1               & 5              & 0                & 3.38E-14             & 1.00E+04               & 8.20E+07                  \\
			RL                & 1               & 5              & 5                & 3.87E-09             & 3.95E+04               & 1.33E+08                  \\
			RL                & 100             & 0              & 0                & 5.12E-53             & 1.00E+02               & 1.76E+07                  \\
			RL                & 100             & 0              & 5                & 3.33E-16             & 2.16E+04               & 8.56E+07                  \\
			RL                & 100             & 5              & 0                & 1.67E-06             & 1.13E+06               & 4.27E+08                  \\
			RL                & 100             & 5              & 5                & \prism T/O       & 2.29E+06               & 5.68E+08                  \\
			RL                & 10000           & 0              & 0                & 4.34E-16             & 7.07E+03               & 4.30E+08                  \\
			RL                & 10000           & 0              & 5                & 1.03E-15             & 1.69E+06               & 5.48E+08                  \\
			RL                & 10000           & 5              & 0                & \prism T/O                     & 3.04E+06               & 1.09E+09                  \\
			RL                & 10000           & 5              & 5                & \prism T/O                     & 6.75E+06               & 1.18E+09                  \\
			RDG               & 1               & 0              & 0                & 1.11E-107            & 1.00E+00               & 3.93E+05                  \\
			RDG               & 1               & 0              & 5                & 1.55E-31             & 1.76E+02               & 5.51E+06                  \\
			RDG               & 1               & 5              & 0                & 6.38E-11             & 1.49E+04               & 9.72E+07                  \\
			RDG               & 1               & 5              & 5                & 6.87E-07             & 5.86E+04               & 1.57E+08                  \\
			RDG               & 100             & 0              & 0                & 5.39E-47             & 8.40E+01               & 1.67E+07                  \\
			RDG               & 100             & 0              & 5                & 2.75E-16             & 2.01E+04               & 7.53E+07                  \\
			RDG               & 100             & 5              & 0                & 1.67E-06             & 1.00E+06               & 4.02E+08                  \\
			RDG               & 100             & 5              & 5                & \prism T/O       & 1.92E+06               & 5.18E+08                  \\
			RDG               & 10000           & 0              & 0                & 4.04E-16             & 4.90E+03               & 4.30E+08                  \\
			RDG               & 10000           & 0              & 5                & 1.02E-15             & 1.67E+06               & 5.45E+08                  \\
			RDG               & 10000           & 5              & 0                & \prism T/O                     & 2.38E+06               & 9.97E+08                  \\
			RDG               & 10000           & 5              & 5                &  \prism T/O                    & 5.39E+06               & 1.06E+09                  \\ \hline
		\end{tabular}%
	}
\end{table}

\begin{table}[p]
	\caption{Enzymatic Futile Cycle}
	\label{tab:efc}
	\resizebox{\textwidth}{!}{%
		\begin{tabular}{lllllll}
			\hline
			\textbf{Approach} & \textbf{Traces} & \textbf{Cycle} & \textbf{Commute} & \textbf{Probability} & \textbf{Time (ms)} & \textbf{Bytes} \\ \hline
			RL                & 1               & 0              & 0                & 1.88E-78             & 1.00E+00               & 1.97E+05                  \\
			RL                & 1               & 0              & 5                & 1.82E-78             & 1.00E+00               & 1.97E+05                  \\
			RL                & 1               & 5              & 0                & 1.72E-07             & 6.00E+00               & 1.97E+05                  \\
			RL                & 1               & 5              & 5                & 1.72E-07             & 3.00E+00               & 1.97E+05                  \\
			RL                & 100             & 0              & 0                & 2.45E-53             & 1.90E+01               & 2.36E+06                  \\
			RL                & 100             & 0              & 5                & 4.73E-48             & 2.80E+01               & 2.95E+06                  \\
			RL                & 100             & 5              & 0                & 1.72E-07             & 2.50E+01               & 3.15E+06                  \\
			RL                & 100             & 5              & 5                & 1.72E-07             & 3.20E+01               & 2.95E+06                  \\
			RL                & 10000           & 0              & 0                & 4.50E-32             & 1.18E+03               & 1.81E+08                  \\
			RL                & 10000           & 0              & 5                & 7.81E-32             & 2.24E+03               & 1.81E+08                  \\
			RL                & 10000           & 5              & 0                & 1.72E-07             & 1.58E+03               & 2.24E+08                  \\
			RL                & 10000           & 5              & 5                & 1.72E-07             & 2.23E+03               & 1.81E+08                  \\
			RDG               & 1               & 0              & 0                & 1.90E-78             & 1.00E+00               & 1.97E+05                  \\
			RDG               & 1               & 0              & 5                & 1.97E-78             & 1.00E+00               & 1.97E+05                  \\
			RDG               & 1               & 5              & 0                & 1.72E-07             & 5.00E+00               & 1.97E+05                  \\
			RDG               & 1               & 5              & 5                & 1.72E-07             & 6.00E+00               & 1.97E+05                  \\
			RDG               & 100             & 0              & 0                & 2.02E-78             & 1.00E+01               & 2.36E+06                  \\
			RDG               & 100             & 0              & 5                & 2.02E-78             & 2.00E+01               & 2.95E+06                  \\
			RDG               & 100             & 5              & 0                & 1.72E-07             & 1.90E+01               & 3.15E+06                  \\
			RDG               & 100             & 5              & 5                & 1.72E-07             & 2.40E+01               & 2.75E+06                  \\
			RDG               & 10000           & 0              & 0                & 2.02E-78             & 6.31E+02               & 1.82E+08                  \\
			RDG               & 10000           & 0              & 5                & 2.02E-78             & 1.70E+03               & 2.24E+08                  \\
			RDG               & 10000           & 5              & 0                & 1.72E-07             & 1.04E+03               & 2.23E+08                  \\
			RDG               & 10000           & 5              & 5                & 1.72E-07             & 1.71E+03               & 2.24E+08                  \\ \hline
		\end{tabular}%
	}
\end{table}

\begin{table}[p]
	\caption{Simplified Motility Regulation}
	\label{tab:smr}
	\resizebox{\textwidth}{!}{%
		\begin{tabular}{lllllll}
			\hline
			\textbf{Approach} & \textbf{Traces} & \textbf{Cycle} & \textbf{Commute} & \textbf{Probability} & \textbf{Time (ms)} & \textbf{Bytes} \\ \hline
			RL                & 1               & 0              & 0                & 2.07E-15             & 0.00E+00               & 1.97E+05                  \\
			RL                & 1               & 0              & 5                & 1.89E-13             & 8.90E+01               & 1.97E+05                  \\
			RL                & 1               & 5              & 0                & 1.23E-07             & 2.74E+05               & 1.97E+05                  \\
			RL                & 1               & 5              & 5                & 2.04E-07             & 2.85E+06               & 1.97E+05                  \\
			RL                & 100             & 0              & 0                & 4.83E-14             & 4.70E+01               & 5.90E+05                  \\
			RL                & 100             & 0              & 5                & 8.57E-11             & 5.89E+03               & 7.08E+06                  \\
			RL                & 100             & 5              & 0                & 1.43E-07             & 7.84E+05               & 2.26E+08                  \\
			RL                & 100             & 5              & 5                & 2.13E-07             & 3.89E+06               & 4.48E+08                  \\
			RL                & 10000           & 0              & 0                & 1.06E-13             & 5.38E+03               & 7.86E+05                  \\
			RL                & 10000           & 0              & 5                & 1.08E-10             & 1.69E+04               & 7.67E+06                  \\
			RL                & 10000           & 5              & 0                & 1.62E-07             & 9.00E+05               & 2.38E+08                  \\
			RL                & 10000           & 5              & 5                & 2.21E-07             & 4.17E+06               & 4.61E+08                  \\
			RDG               & 1               & 0              & 0                & 2.07E-15             & 0.00E+00               & 1.97E+05                  \\
			RDG               & 1               & 0              & 5                & 1.89E-13             & 8.70E+01               & 1.97E+05                  \\
			RDG               & 1               & 5              & 0                & 1.23E-07             & 3.78E+05               & 1.97E+05                  \\
			RDG               & 1               & 5              & 5                & 2.04E-07             & 2.84E+06               & 1.97E+05                  \\
			RDG               & 100             & 0              & 0                & 2.07E-15             & 7.00E+00               & 1.97E+05                  \\
			RDG               & 100             & 0              & 5                & 1.89E-13             & 8.80E+01               & 1.57E+06                  \\
			RDG               & 100             & 5              & 0                & 1.23E-07             & 3.69E+05               & 1.58E+08                  \\
			RDG               & 100             & 5              & 5                & 2.04E-07             & 2.85E+06               & 3.78E+08                  \\
			RDG               & 10000           & 0              & 0                & 2.07E-15             & 3.06E+02               & 7.86E+05                  \\
			RDG               & 10000           & 0              & 5                & 1.89E-13             & 3.91E+02               & 1.57E+06                  \\
			RDG               & 10000           & 5              & 0                & 1.23E-07             & 3.73E+05               & 1.59E+08                  \\
			RDG               & 10000           & 5              & 5                & 2.04E-07             & 2.85E+06               & 3.78E+08                  \\ \hline
		\end{tabular}%
	}
\end{table}

\begin{table}[p]
	\caption{Single Species Production Degradation}
	\label{tab:sspd}
	\resizebox{\textwidth}{!}{%
		\begin{tabular}{lllllll}
			\hline
			\textbf{Approach} & \textbf{Traces} & \textbf{Cycle} & \textbf{Commute} & \textbf{Probability} & \textbf{Time (ms)} & \textbf{Bytes} \\ \hline
RL                & 1               & 0              & 0                & 1.94E-16             & 0.00E+00               & 1.97E+05                  \\
RL                & 1               & 0              & 5                & 2.67E-07             & 1.00E+00               & 1.97E+05                  \\
RL                & 1               & 5              & 0                & 2.23E-07             & 0.00E+00               & 1.97E+05                  \\
RL                & 1               & 5              & 5                & 2.92E-07             & 3.00E+00               & 3.93E+05                  \\
RL                & 100             & 0              & 0                & 4.97E-13             & 4.50E+01               & 3.93E+05                  \\
RL                & 100             & 0              & 5                & 2.67E-07             & 5.60E+01               & 5.90E+05                  \\
RL                & 100             & 5              & 0                & 2.49E-07             & 3.40E+01               & 5.90E+05                  \\
RL                & 100             & 5              & 5                & 2.95E-07             & 6.00E+01               & 5.90E+05                  \\
RL                & 10000           & 0              & 0                & 1.40E-07             & 3.23E+03               & 7.86E+05                  \\
RL                & 10000           & 0              & 5                & 2.79E-07             & 3.27E+03               & 7.86E+05                  \\
RL                & 10000           & 5              & 0                & 2.49E-07             & 3.25E+03               & 7.86E+05                  \\
RL                & 10000           & 5              & 5                & 2.95E-07             & 3.30E+03               & 5.90E+05                  \\
RDG               & 1               & 0              & 0                & 1.94E-16             & 0.00E+00               & 1.97E+05                  \\
RDG               & 1               & 0              & 5                & 2.67E-07             & 1.00E+00               & 1.97E+05                  \\
RDG               & 1               & 5              & 0                & 2.23E-07             & 1.00E+00               & 1.97E+05                  \\
RDG               & 1               & 5              & 5                & 2.92E-07             & 3.00E+00               & 3.93E+05                  \\
RDG               & 100             & 0              & 0                & 1.94E-16             & 1.00E+01               & 1.97E+05                  \\
RDG               & 100             & 0              & 5                & 2.67E-07             & 1.10E+01               & 1.97E+05                  \\
RDG               & 100             & 5              & 0                & 2.23E-07             & 1.00E+01               & 1.97E+05                  \\
RDG               & 100             & 5              & 5                & 2.92E-07             & 1.10E+01               & 3.93E+05                  \\
RDG               & 10000           & 0              & 0                & 1.94E-16             & 8.10E+02               & 1.97E+05                  \\
RDG               & 10000           & 0              & 5                & 2.67E-07             & 8.86E+02               & 1.97E+05                  \\
RDG               & 10000           & 5              & 0                & 2.23E-07             & 8.09E+02               & 1.97E+05                  \\
RDG               & 10000           & 5              & 5                & 2.92E-07             & 8.46E+02               & 1.97E+05                  \\ \hline
\end{tabular}
}
\end{table}

\newpage

\section{CRN \& CTSVAS Input Format} 
\label{appendix:language}

\ragtimer relies on syntax and semantics from a custom input format that is designed to be straightforward for users without an in-depth understanding of CTMC modeling specifications. This appendix gives an overview of this input format.

\subsection{Overview}

The input format is designed to be minimalistic and simple, reducing the burden of education for non-technical users, including synthetic biologists and experts in various domains. Each line of the input specifies a unique property of the model, with a flexible dictionary of keywords in plain English.

For example, a CTSVAS input follows the format Listing~\ref{lst:general_vas}. The first 3 lines declare the names of variables along with their initialization. Line 4 declares a target variable and its desired value. To reach the target, this model 
must increase \texttt{S3} from \texttt{300} to \texttt{340}. Lines 5-8 and 9-12 declare two transitions. Lines 5 and 9 declare the transition names. Lines beginning with \texttt{increase} and \texttt{decrease} define the update vector and enabled bounds for each transition. Line 8 and 12 declare transition rate constants, adhering to the CRN Stochastic Chemical Kinetics assumption and probability semantics~\cite{Myers_Engineeringgeneticcircuits_2010}.

\begin{figure}[h]
	\centering
	\begin{lstlisting}[style=vas]
var S1 init 100
var S2 init 200
var S3 init 300
target S3 = 340
transition R1
	decrease S1 10
	increase S2 3
	const 0.4
transition R2
	decrease S1 1
	increase S3 1
	const 0.6
	\end{lstlisting}
	\caption{Custom CTSVAS Input Format Example}
	\label{lst:general_vas}
\end{figure}

\subsection{Syntax \& Semantics}

This section formally defines the syntax and semantics of the input format for the parsing and construction of a generic CTSVAS. Declarations occupy exactly one line each. Tokens are space- or tab-separated, and newline characters are disallowed within a variable declaration.

\subsubsection{Variables and Initial States.}
For user convenience, a CTSVAS may include a vector of variable names and a total ordering $\leqslant$ imposed thereon and defined by the order in which variables are declared in the input file. The total order $\leqslant$ is preserved through the entire model. Variable declarations traditionally appear at the beginning of the input specification, but they may be present on any line. Variables must be declared exactly once per model.

Any line whose first token is in $\{$ \texttt{var}, \texttt{variable}, \texttt{species} $\}$ is interpreted as a variable declaration. Within a variable declaration, the second token is interpreted as the variable's name (used only for user convenience). If a third token is present, it must be in $\{$\texttt{init}, \texttt{initial}$\}$, and its purpose is to declare the initial state for that variable. The fourth token must be present if the third token is present, and it defines the initial state corresponding to the named variable. If only two tokens are present, the variable is assigned an initial value of 0.

For example, Listing~\ref{lst:variables} shows a simple variable initialization. The order imposed on these variables is \texttt{S1} $\leqslant$ \texttt{S2} $\leqslant$ \texttt{S3}. The variable name vector is $[\texttt{S1}, \texttt{S2}, \texttt{S3}]$, and $\initialState = [100, 200, 300]$. 

\begin{figure}[htbp]
	\centering
	\begin{lstlisting}[style=vas]
var S1 init 100
var S2 init 200
var S3 init 300
	\end{lstlisting}
	\caption{Variable Declaration}
	\label{lst:variables}
\end{figure}

\subsubsection{Transitions.}
For user convenience, transitions are assigned names, and their corresponding vectors use the total ordering $\leqslant$ imposed during variable parsing. Any line whose first token is in $\{$\texttt{transition}, \texttt{reaction}$\}$ is considered a transition declaration. The second token of the line is the transition's name.

Consider $\vec{x}[v]$ to indicate a valuation of vector $\vec{x}$ at the position of the variable named $v$. Consider a line $l$ such that transition $\transition[\alpha]$ is declared at the greatest line number not exceeding the line number of $l$. If the first token of $l$ is in $\{$\texttt{decrease}, \texttt{decrement}, \texttt{consume}, \texttt{increase}, \texttt{increment}, \texttt{produce}$\}$, this defines a deviation from the default values for $\update{\alpha}$ and $\guard{\alpha}$. Contradictory updates to variables (i.e., multiple increases or decreases for a single variable) are not permitted.

The first token indicates whether the variable should increase or decrease upon execution of $\transition[\alpha]$. A token in $\{$\texttt{increase}, \texttt{increment}, \texttt{produce}$\}$ indicates an increase, while a token in $\{$\texttt{decrease}, \texttt{decrement}, \texttt{consume}$\}$ indicates a decrease.
Within $l$, the second token $l_2$ indicates the affected variable $v$, and the third token ($i_3$ for increasing and $d_3$ for decreasing), indicates the amount by which transition $\transition[\alpha]$ decreases the affected variable. If only two tokens are present, the value of $i_3$ or $d_3$ is assumed to be $1$. Formally, for each line $l$ describing variable $v$, $\update{\alpha}[v] = \sum_{\forall l} i_3-d_3 $ and $\enabled{i}[v] = \sum_{\forall l} d_3 $. Intuitively, the update vector for each transition is the sum of increases and decreases for each variable, and the enabled vector is the sum of decreases for each variable.

Similarly, if the first token of $l_t$ is in $\{$\texttt{const}, \texttt{rate}$\}$, the probability function reflects a rate constant matching the second token of $l_t$. If a rate formula is desired instead, the token \texttt{formula} should precede the specification of a rate formula using mathematical notation identical to expressions specified by the Rust language. For example, if the rate should equal 3.25 times the value of v, one may specify \texttt{formula 3.25 * v}.

For example, Listing~\ref{lst:transitions} shows a simple transition description: the update vector matches the specification as $\update{1} = [-10, 3, 0]$, and the guard $\guard{1} = [10, 0, 0]$. Similarly, $\update{2} = [-1, 0, 1]$ and $\guard{2} = [1, 0, 0]$. The rate constant for \texttt{R1} is $0.4$, and the rate constant for \texttt{R2} is $0.6$.

\begin{figure}[htbp]
	\centering
	\begin{lstlisting}[style=vas]
transition R1
	decrease S1 10
	increase S2 3
	const 0.4
transition R2
	decrease S1 1
	increase S3 1
	const 0.6
	\end{lstlisting}
	\caption{Transition Specification}
	\label{lst:transitions}
\end{figure}

\subsubsection{Property Specification.}
A target property (often a violation of a safety property) is specified on lines whose first token is in $\{$\texttt{target}, \texttt{goal}, \texttt{prop}, \texttt{check}$\}$. This token is followed by an expression that uses mathematical functions defined in the Rust specification. At present, exactly one target should be present in a model. For example, Listing~\ref{lst:property} describes a target in which the value of \texttt{S3} is $340$.

\begin{figure}[htbp]
	\centering
	\begin{lstlisting}[style=vas]
target S3 = 340
	\end{lstlisting}
	\caption{Property Specification}
	\label{lst:property}
\end{figure}

\subsection{Example: MYP Model}
Listing~\ref{lst:mypinput} shows the Modified Yeast Polarization model described in Section~\ref{sec:motivating} encoded in the custom input format. 

\begin{figure}[htbp]
	\centering
	\begin{lstlisting}[style=vas]
species R init 50
species L init 2
species RL init 0
species G init 50
species GA init 0
species GBG init 0
species GD init 0
target GBG = 50
reaction R1
	produce R 1
	const 0.0038
reaction R2
	consume R 1
	const 0.0004
reaction R3
	consume R 1
	consume L 1
	produce RL 1
	produce L 1
	const 0.042
reaction R4
	consume RL 1
	produce R 1
	const 0.010
reaction R5
	consume RL 1
	consume G 1
	produce GA 1
	produce GBG 1
	const 0.011
reaction R6
	consume GA 1
	produce GD 1
	const 0.100
reaction R7
	consume GBG 1
	consume GD 1
	produce G 1
	const 1050
reaction R8
	produce RL 1
	const 3.210
	\end{lstlisting}
	\caption{Custom CTSVAS Input Format for MYP model}
	\label{lst:mypinput}
\end{figure}

\end{subappendices}

\end{document}